# Extraordinary electrical conductance through amorphous non-conducting polymers under vibrational strong coupling

Sunil Kumar[1#], Subha Biswas[1#], Umar Rashid[1], Kavya S. Mony[1], Gokul Chandrasekharan[1], Francesco Mattiotti[2], Robrecht M. A. Vergauwe[3], David Hagenmuller[2*], Veerabhadrarao Kaliginedi[1*], Anoop Thomas[1*]

[1]Inorganic and Physical Chemistry, Indian Institute of Science, Bengaluru, India.
[2]University of Strasbourg and CNRS, CESQ and ISIS (UMR 7006), France.
[3]Nanoscience Center and Department of Chemistry, University of Jyväskylä, Finland.
*Correspondence to: dhagenmuller@unistra.fr; vkaliginedi@iisc.ac.in; athomas@iisc.ac.in
#These authors contributed equally

**Abstract**

Achieving electrical conductance in amorphous non-doped polymers is a challenging task. Here, we show that vibrational strong coupling of the aromatic C-H(D) out-of-plane bending modes of polystyrene, deuterated polystyrene, and poly (benzyl methacrylate) to the vacuum electromagnetic field of the cavity enhance the electrical conductivity by at least six orders of magnitude compared to the uncoupled polymers. The conductance is thermally activated at the onset of strong coupling. It becomes temperature and cavity path length independent at the highest coupling strengths, giving rise to the extraordinary electrical conductance in these polymers. The electrical characterizations are performed without external light excitation, demonstrating the role of quantum light in enhancing the long-range coherent transport even in amorphous non-conducting polymers.

**Main Text**

The exciting aspect of quantum light-matter interaction is that the vacuum fluctuations of the electromagnetic radiation field can strongly couple with the molecular transition dipole to form hybrid light-matter states known as polaritonic states [1,2]. Ultrafast imaging studies under electronic strong coupling (ESC) show that polaritons diffuse faster and can reach the ballistic regime above a critical photon-exciton fraction [3,4], which is likely a reason for the enhanced transport properties of matter under ESC [5–13]. Vibrational strong coupling (VSC) produces polaritonic states of molecular vibrational modes and light [14,15], which has notably impacted chemical reactivity [16–26] and intermolecular interactions [27–30]. However, the role of VSC in organic electronics is unexplored as it might be less intuitive to think that the VSC of the ground-state vibrational mode would affect the electrical conductance.

Interestingly, mode-selective vibrational excitation with lasers generates transient conductance modulation in molecular systems [31,32]. Vibrational strong coupling should provide an alternative to the former short-lived process since vibro-polaritonic states can form at ambient conditions without external light excitation [1]. To check the hypothesis, we designed experiments to measure electrical conductance through organic molecules under VSC. While optimizing the Fabry-Perot cavity to suit our electrical measurement system, we fortuitously observed extraordinary electrical conductance through a cavity containing strongly coupled polystyrene (PS). Our extensive investigation using PS, deuterated PS (PS-$d_8$), poly(benzyl methacrylate) (PBMA) and poly(methyl methacrylate) (PMMA) shows that VSC of the aromatic C-H(D) out-of-plane bending

($\delta$(Ar. CH(D))) vibrational modes (Fig. S1) enhance the electrical conductance by c.a. six orders of magnitude compared to uncoupled (non-cavity) polymers.

The Fabry-Perot cavities in our experiments constitute a bottom and top Au mirror (10 nm) separated by the active polymer film, supported on an intrinsic Si as shown schematically in Fig.1A. We varied the cavity path length (polymer thickness) across samples (1 to 9 µm) to obtain the resonance between the polymer vibrational transition ($\hbar\omega_m$) and one of the cavity modes ($\hbar\omega_c$). For electrical conductance measurements, we used a home-built setup (Fig S2) with a liquid metal probe (eutectic mixture of gallium and indium; EGaIn) to make a non-deforming soft contact with the cavity top mirror (Fig. 1A), which is otherwise challenging with a solid metal probe [33]. The EGaIn-based eutectic liquid is widely used to analyze the electrical properties of self-assembled molecular layers [34,35]. To examine the vertical electrical conductance, we collect the I-V response (Fig. S3, S4) within a bias voltage window of +1 to -1 V and obtain a statistically significant master curve of each cavity by measuring at least six different electrical junctions (720 sweeps) and plot the mirror area normalized current density |J| against voltage V [36]. Polymer spin coating can sometimes create pathlength inhomogeneity in large area samples (2 x 2 cm); typically, the sides may have thickness heterogeneity and no cavity modes (Fig.S5). In a complementary experiment, we prepared cross-mirror geometry samples, a configuration commonly used in organic electronic devices [33], and limited the cavity only to the homogenous region of the film (5 x 5 mm). Measurements at the cross-mirror cavity (Fig S6) and the regular cavity's homogeneous region resulted in identical spectral and J-V characteristics. The spectral and electrical characterization data presented here correspond to the measurements done in the homogeneous region of the cavity. Refer to Supporting Information for sample preparation and characterization details.

We begin our analysis of the effect of VSC on the electrical conductance of PS by focusing on its $\delta$(Ar. CH) vibration with the peak at 698 cm$^{-1}$ (Fig. 1B, black curve) [37]. The out-of-plane bending vibrations typically facilitate vibronic coupling [38], a condition aiding the transient conductance enhancement under vibrational mode selective excitation [31,32]. The red curve in Fig. 1B shows the infrared transmission spectrum of the strongly coupled PS, where the cavity second mode ($\hbar\omega_c$ = 698 cm$^{-1}$) is on-resonance with the $\delta$(Ar. CH) transition. The two polariton bands (P+ and P-) show the characteristic spectral signature of light-matter strong coupling [1]. The polaritonic peaks are equally separated from the $\hbar\omega_m$, with a Rabi splitting energy ($\hbar\Omega_R$) of 53 cm$^{-1}$, which is larger than the FWHM of both the cavity mode (20 cm$^{-1}$) and the PS $\delta$(Ar. CH) vibration (16 cm$^{-1}$), indicating that the system is under VSC. The clear anti-crossing of the polariton peaks (Fig.1C black diamonds) and its excellent correlation with the transfer-matrix simulated [14] angle-dependent dispersion plot (Fig.1C contour plot; Fig.S7) confirm the formation of the hybrid light-matter states.

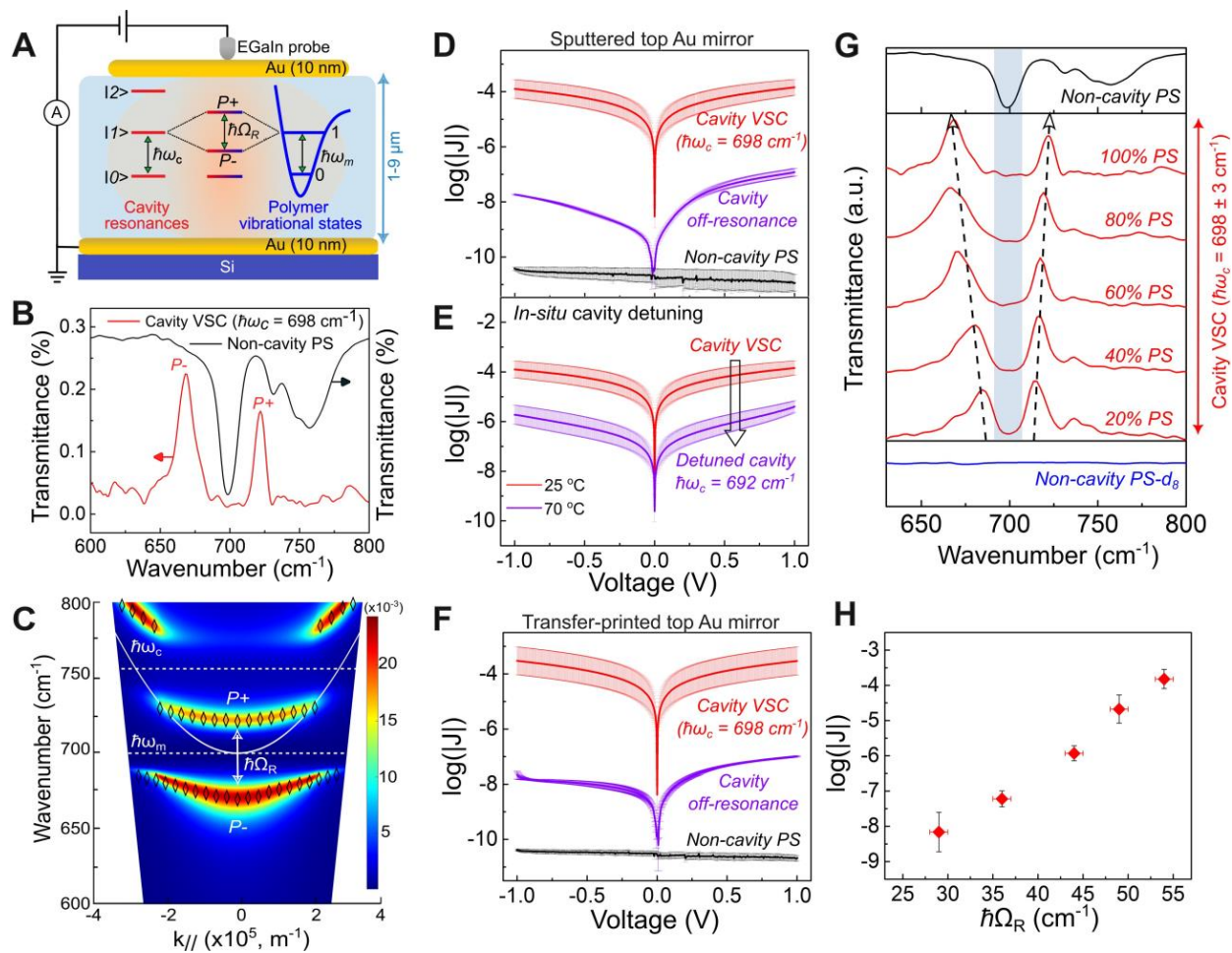

**Figure 1. Vibrational strong coupling of $\delta$(Ar. CH) transition of PS and electrical conductance**. (A) Schematic illustration of the VSC, the formation of the polaritonic states (P+ and P-) and the electrical conductance measurement scheme. (B) Infrared transmission spectrum of the cavity under VSC of the $\delta$(Ar. CH) transition of PS (red curve) and a thin film of PS (black curve). (C) The angle-dependent dispersion of the polaritonic peaks plotted against the in-plane wave vector (black diamonds) overlayed on the simulated dispersion contour of the strongly coupled PS. The solid and dashed white curve represent the simulated dispersion of the uncoupled cavity mode ($\hbar\omega_c$= 698 cm$^{-1}$), and the $\delta$(Ar. CH) transitions of PS ($\hbar\omega_m$ = 698 cm$^{-1}$ and 758 cm$^{-1}$), respectively. (D) Shows the plots of J-V curves under cavity VSC of the $\delta$(Ar. CH) transition of PS (red curve, $\hbar\omega_m$= $\hbar\omega_c$= 698 cm$^{-1}$), under off-resonance (purple curve), and non-cavity (black curve). (E) Shows the reduction in the current density (purple curve) upon *in-situ* detuning of the strongly coupled cavity by heating to 70 °C. (F) The J-V curves of transfer printed top Au mirror cavities under cavity VSC (red curve), off-resonance (purple curve), and non-cavity (black curve). The non-cavity refers to a top mirrorless region of the on-resonance sample. The thick solid line in D, E, and F represents the statistically significant master curve extracted from 720 sweeps, and the thin lines show the standard deviation. (G) The IR transmission spectra of the PS under cavity VSC (red curves, $\hbar\omega_m$= 698 cm$^{-1}$; $\hbar\omega_c$= 698±3 cm$^{-1}$) correspond to various PS weight percentages mentioned in the figure, non-cavity PS (black curve) and PS-$d_8$ (blue curve), respectively. The shaded region indicates the FWHM of the PS $\delta$(Ar. CH) transition. (H) Variation of current density as a function of $\hbar\Omega_R$.

The VSC of the $\delta$(Ar. CH) of PS ($\hbar\omega_m = \hbar\omega_c =$ 698 cm$^{-1}$) showed a surprisingly high current density, ~1 x10$^{-4}$ Acm$^{-2}$ at 1V, which is six and four orders of magnitude higher than the non-cavity PS and off-resonance cavity ($\hbar\omega_m \neq \hbar\omega_c$), respectively (Fig.1D, Fig. S8). The non-cavity PS is the reference sample in which PS is spin-coated on top of Au supported on Si (Fig.S9). It may have a lower current density [39] than measured here, as it is the noise floor of our measurement system (Fig. S9). Subsequently, we detuned the cavity under VSC *in situ* by heating (70 °C). Slight thermal expansion of polymer increases the cavity path length and redshifts the $\hbar\omega_c$ to 692 cm$^{-1}$, detuning from the resonance. As a result, the current density drops two orders of magnitude (Fig. 1E purple curve). Cooling the cavity to 25 °C recovers the on-resonance and, therefore, the enhanced current density (Fig. S10), indicating the apparent correlation between the cavity VSC and extraordinary conductance. Further, we made a series of on- and off-resonance PS cavities by transfer printing [40] the top Au film. The J-V curves (Fig. 1F and Fig.S11-S20) of transfer printed cavities resemble those prepared by direct sputtering, confirming the cavity VSC origin of the extraordinary conductance and ruling out any top mirror sputtering artifacts. The X-ray photoelectron spectroscopic depth profile analysis of cavities (Fig S21) verified the absence of Au penetration into the PS during top mirror sputtering, which agrees with the X-ray reflectivity study [41]. The observed extraordinary conductance under cavity VSC of the $\delta$(Ar. CH) mode is independent of the molecular weight (Mw = 35000, 192000, and 292000 Da) and the commercial origin of PS (Fig.S22 and S23). Refer to Supporting Information for details.

To ascertain the VSC effect, we carried out experiments using PS-$d_8$. The C-D vibrational modes are redshifted due to the higher reduced mass compared to the C-H, therefore the $\delta$(Ar. CD) transition of PS-$d_8$ is at 466 cm$^{-1}$ (Fig.S1), resulting in a vibrational mode-free window around 698 cm$^{-1}$ (Fig.1G, blue curve) [42]. Firstly, we analyzed the current density of a PS-$d_8$-filled cavity ($\hbar\omega_c$ = 700 cm$^{-1}$) and confirmed that a cavity mode without VSC does not induce extraordinary conductance (Fig. S24). Secondly, we investigated the effect of strong coupling strength, given by the $\hbar\Omega_R$, on the electrical conductance of PS. To perform these experiments, we prepared cavities with varying PS concentrations by mixing with PS-$d_8$ (Mw = 255000 Da) and ensuring the resonance between cavity first mode and $\delta$(Ar. CH) of PS ($\hbar\omega_m = \hbar\omega_c =$ 698 cm$^{-1}$). Since PS-$d_8$ does not absorb in the region of interest (Fig.1G), the observed $\hbar\Omega_R$ is only due to the cavity VSC of the PS. The $\hbar\Omega_R$ varies from 54±1 cm$^{-1}$ to 29±1 cm$^{-1}$ when the PS wt.% in the thin film decreases from 100% to 20 % (Fig.1G red curves). At the lowest PS concentration, the $\hbar\Omega_R$ (29±1 cm$^{-1}$) is larger than the FWHM (16 cm$^{-1}$) of the $\delta$(Ar. CH) but slightly greater than or equal to the $\hbar\omega_c$ FWHM (26±3 cm$^{-1}$) indicating the onset of VSC. The linear dependence of $\hbar\Omega_R$ on the square root of PS concentration (Fig.S25) further confirms the VSC of the $\delta$(Ar. CH) in all the cavities. The measured current density scales with $\hbar\Omega_R$, proving that the extraordinary enhancement in conductance of PS is indeed a cavity VSC-induced phenomenon (Fig.1H).

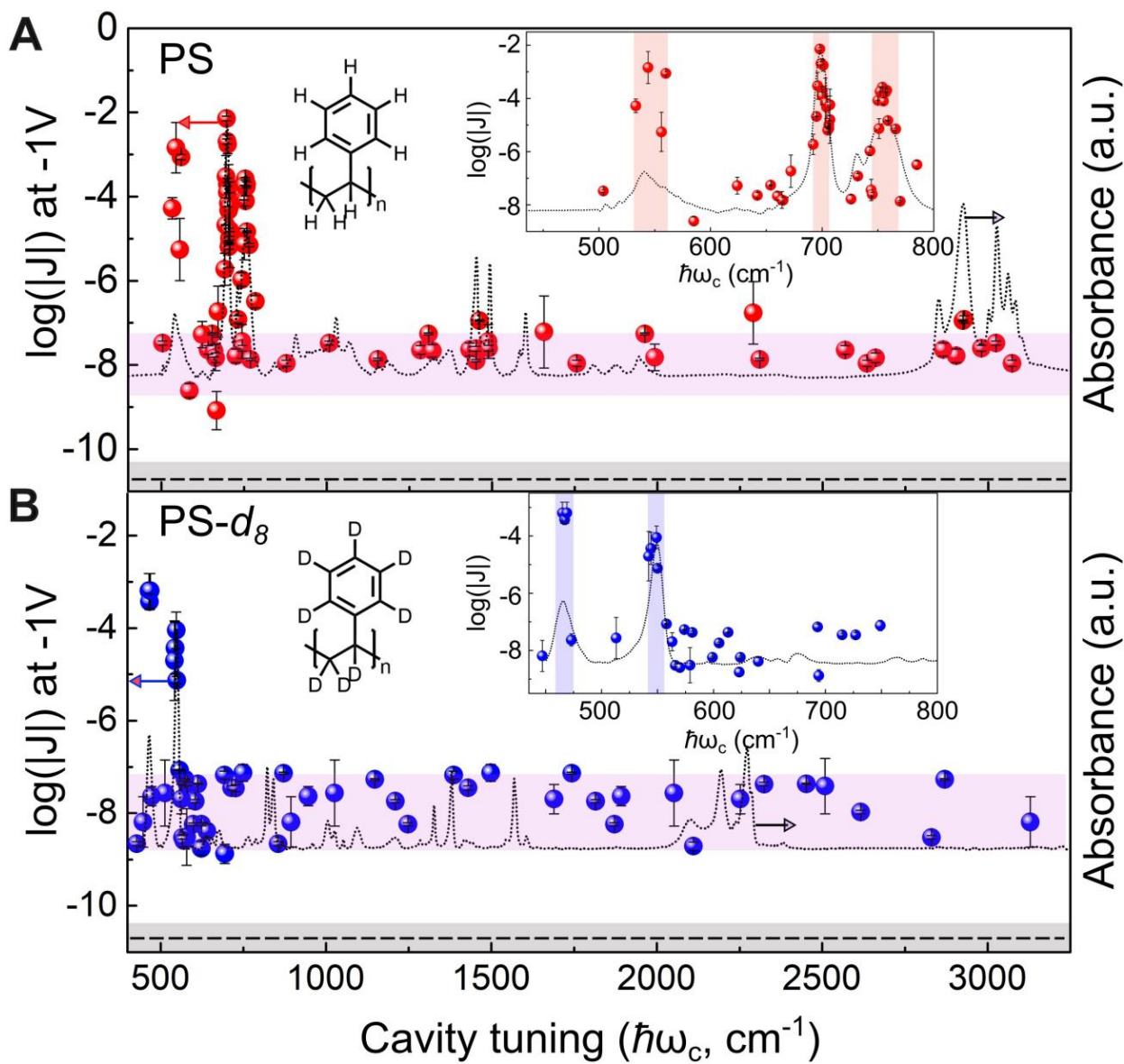

**Figure 2. Electrical conductance through PS and PS-$d_8$ as a function of cavity tuning.** (A,B) Cavity action spectrum showing the impact of cavity VSC of different vibrational modes on electrical conductance through PS (red spheres) and PS-$d_8$ (blue spheres), respectively. The spheres indicate the current density observed at -1V for the cavity where one of its modes resonates at the wavenumber indicated in the x-axis. The purple-shaded region highlights the conductance of the cavities, which are off-resonance or strongly coupled to the non-$\delta$(Ar. CH) and non-$\delta$(Ar. CD) vibrational transition of PS and PS-$d_8$, respectively. The black dashed line shows the measured current density value of non-cavity PS and PS-$d_8$ with standard deviation shaded in grey. The black dotted line shows the infrared absorption of PS and PS-$d_8$. The inset shows the zoom corresponding to the $\delta$(Ar. CH) and $\delta$(Ar. CD) vibrational transition region of PS and PS-$d_8$, respectively. The inset's red- and blue-shaded area shows the FWHM of each $\delta$(Ar. CH) and $\delta$(Ar. CD) vibrational mode of PS and PS-$d_8$, respectively.

Besides the $\delta$(Ar. CH) at 698 cm$^{-1}$, the atactic PS exhibits other strong vibrational transitions that can be strongly coupled (Fig. 2A, black dashed curve). It includes the additional $\delta$(Ar. CH) at 541 cm$^{-1}$ and 758 cm$^{-1}$, deformation of CH$_2$ (1452 cm$^{-1}$), the ring in-plane stretching (1492 cm$^{-1}$) and

C-H asymmetrical stretching (2924 cm$^{-1}$) [37]. Subsequently, we prepared cavities strongly coupled to the other vibrational modes of PS to check if any mode selectivity exists for extraordinary conductance. The action spectrum (Fig. 2A) acquired from 50 different cavities shows that the VSC of only the three $\delta$(Ar. CH) transitions of PS ($\hbar\omega_m$ = 698 cm$^{-1}$, 758 cm$^{-1}$, and 541 cm$^{-1}$) induce extraordinary conductance enhancement. The cavity VSC of other PS vibrations (1452 cm$^{-1}$, 1492 cm$^{-1}$, and 2924 cm$^{-1}$) yields current density in the range of 10$^{-8}$ to 10$^{-7}$ Acm$^{-2}$ (purple-shaded region in Fig. 2A, Table S2, S3), similar to off-resonance cavities. Further, we analyzed the strongly coupled PS-$d_8$ cavities to benchmark the correlation between the VSC of the ring out-of-plane bending modes and extraordinary conductance. Interestingly, the cavity VSC of the two $\delta$(Ar. CD) transitions of PS-$d_8$ (466 cm$^{-1}$ and 549 cm$^{-1}$) shows extraordinary electrical conductance, with quantitative similarities to PS under cavity VSC of the $\delta$(Ar. CH) transitions (698 cm$^{-1}$ and 758 cm$^{-1}$), indicating the remarkable vibrational mode selectivity of VSC (Fig. 2B, Fig. S27 and Table S2,S4). The frequency of the third $\delta$(Ar. CD) of PS-$d_8$ could be below our spectrometers detector range and is not detected (Fig.2B). Akin to PS, the off-resonance PS-$d_8$ cavities and those under the cavity VSC of non- $\delta$(Ar. CD) modes showed similar conductance values (purple shaded region in Fig.2B), indicating a common origin. Supporting Information provide the individual spectral and electrical characterization data.

Interestingly, we observe similar J-V curves and current density values (c.a. 10$^{-7}$ Acm$^{-2}$ at 1V) for the cavity VSC of $\delta$(Ar. CH) with the lower $\hbar\Omega_R$, cavity VSC of non-$\delta$(Ar. CH(D)) modes, and off-resonance cavities. The common factor here could be the strong coupling of surface plasmon to the mirror $\delta$(Ar. CH(D)) vibrational dipoles of PS [43] and PS-$d_8$. In an off-resonance cavity, polymer vibrational dipoles could indeed couple to the near field of the two Au mirrors (i.e. to two overlapping surface plasmons) [44], resulting in a weak but non-negligible current enhancement, similar to the cavity VSC of the $\delta$(Ar. CH) vibration of PS with the lower $\hbar\Omega_R$. The absence of conductance enhancement in a non-cavity (Fig.1) could be due to the weaker field confinement of Au mid-IR plasmons when only the bottom mirror is present [44]. The absence of extraordinary conductance under the cavity VSC of non-$\delta$(Ar. CH(D)) modes indicates that surface plasmon-non-$\delta$(Ar. CH(D)) strong coupling has minimal contribution towards transport.

Polystyrene and PS-$d_8$ can be assigned to the $C_{2V}$ point group in localized symmetry terms, treating it as monosubstituted benzene [37]. Then, the aromatic out-of-plane bending modes belong to $B_2$ symmetry and have the transition moments perpendicular to the plane of the benzene ring, while other aromatic vibrational modes are in the plane of the ring with $B_1$ and $A_1$ symmetry [37]. In the present scenario, the $\delta$(Ar. CH(D)) modes could favor vibronic mixing and facilitate the interactions between the benzene rings across different layers [45], leading to the extraordinary enhancement in conductance. In contrast, the in-plane vibrations do not favor the interlayer interactions and vibronic coupling [38].

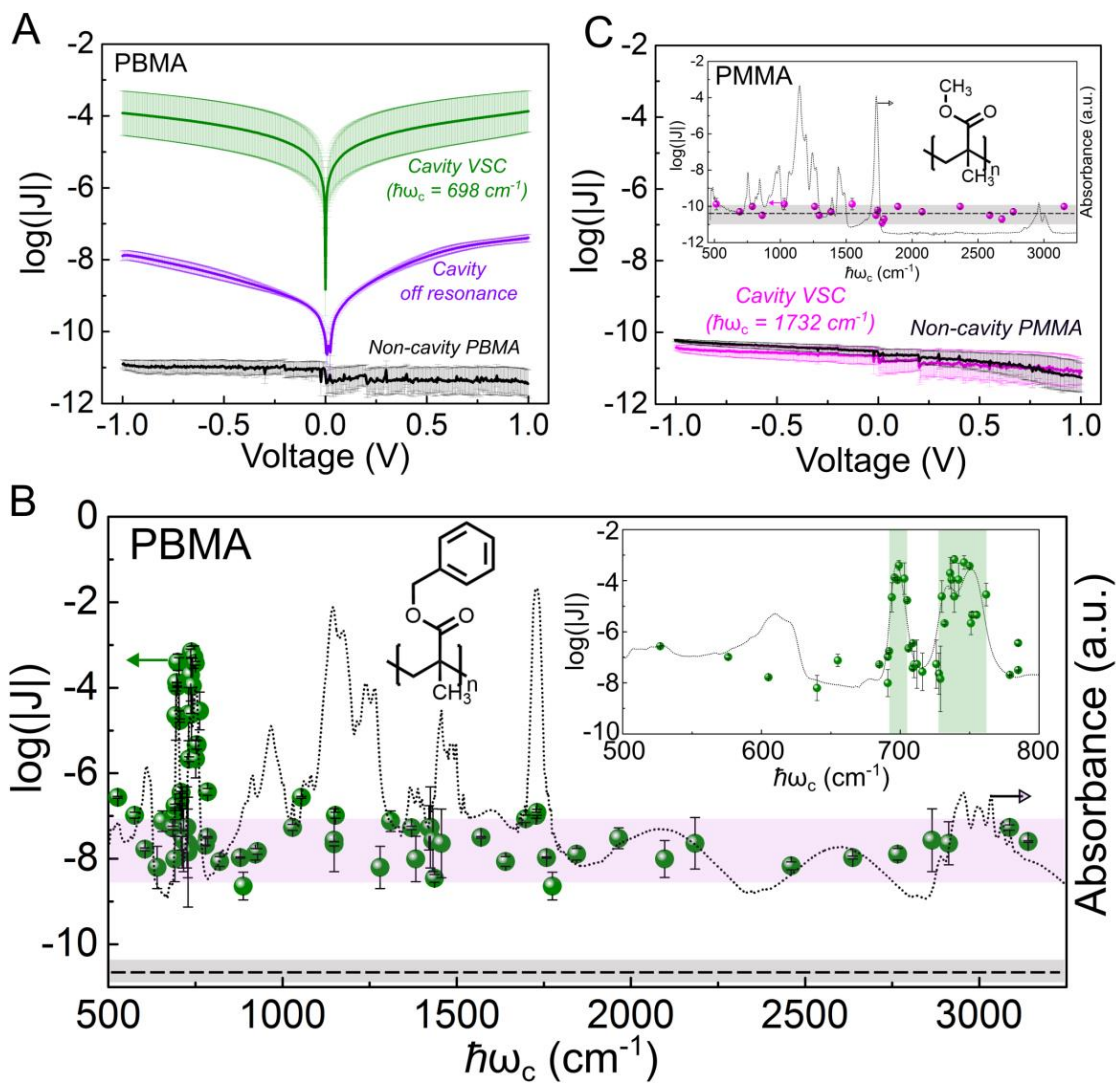

**Figure 3. Extraordinary electrical conductance through PBMA**. (A) The J-V curve corresponds to PBMA under cavity VSC of its $\delta$(Ar. CH) transition ($\hbar\omega_m = 698$ cm$^{-1}$; green curve), cavity off-resonance (purple curve), and non-cavity (black curve). (B) The cavity action spectrum shows the impact of cavity VSC of different vibrational modes on electrical conductance through PBMA. The green spheres indicate the current density observed at -1V for the cavity where one of its modes resonates at the wavenumber indicated in the x-axis. The purple-shaded region highlights the conductance of cavities, which are off-resonance or strongly coupled with the non-$\delta$(Ar. CH) PBMA vibrational transitions. The black dashed line shows the measured current density value of non-cavity PBMA with standard deviation shaded in grey. The black dotted curve shows the IR absorption of PBMA. The inset shows the zoom region corresponding to the $\delta$(Ar. CH) vibrational transition of PBMA, and the green-shaded area shows the FWHM of each $\delta$(Ar. CH) vibrational mode. (C) The J-V curve of PMMA under cavity VSC of its C=O stretching vibrational transition ($\hbar\omega_m = 1729$ cm$^{-1}$; pink curve) and non-cavity (black curve). The inset shows the impact of cavity VSC of different vibrational modes on electrical conductance through PMMA. The pink spheres indicate the current density observed at -1V for the cavity where one of its modes resonates at the wavenumber indicated in the x-axis.

To verify the generality of the above hypothesis, we analyzed the conductance through PBMA (Fig.3B), a polymer possessing $\delta$(Ar. CH) modes, and PMMA (Fig. 3C), an aliphatic polymer having no ring vibrations. Interestingly, we observe the extraordinary conductance enhancement through PBMA under the cavity VSC of its $\delta$(Ar. CH) vibrations, 698 cm$^{-1}$ and 751 cm$^{-1}$ (Fig.3A, B and Fig.S28) and improved current density for cavities which are off-resonance or strongly coupled to non-$\delta$(Ar. CH) transitions (Fig.3B) due to the surface plasmon strong coupling with the $\delta$(Ar. CH) transitions. We confirmed the role of cavity VSC by tuning and detuning the resonance condition through *in-situ* heating and cooling experiments (Fig. S29). In contrast, there is no measurable current through PMMA under the cavity VSC of its different vibrational modes (Fig. 3C, Fig.S30). It ascertains the need for $\delta$(Ar.CH(D)) transition to facilitate the vibronic coupling and interlayer interaction. Remarkably, these observations point to a general phenomenon where VSC can be a tool to tune the conductance properties of non-conducting polymers.

In Fig. 4A, we plot the dependence of the measured current density of PS, PS-$d_8$, PBMA and PMMA against polymer thickness (cavity pathlength). The extraordinary conductance enhancement only depends on the cavity VSC of the $\delta$(Ar. CH(D)) vibrations and is independent of cavity mode number and, therefore, the polymer film thickness. Additionally, the absence of any detectable current through PMMA indicates that the Au mirrors do not contribute to the electrical transport as an impurity. Refer to Supporting Information for the spectral and electrical analysis data of all the samples.

To better understand the nature of the transport in our system, we analyzed the temperature dependence of PS conductance ($\hbar\omega_m$ = 698 cm$^{-1}$) under cavity VSC with varying $\hbar\Omega_R$, in a range of 25 ºC to 50 ºC by maintaining $\hbar\omega_c$ within the limit of 698±3 cm$^{-1}$ (Fig.4B, Fig.S31). Remarkably, for the cavities with larger $\hbar\Omega_R$ (100% and 80% PS), the current density is independent of temperature (Fig.4B). In contrast, cavities with lower $\hbar\Omega_R$ (40% and 20% PS) showed a thermal activation (Fig.4B). We further analyzed the temperature dependence of PS, PS-$d_8$, and PBMA under the cavity VSC of its $\delta$(Ar. CH(D)) mode ($\hbar\omega_m$=758 cm$^{-1}$ (PS), 751 cm$^{-1}$ (PBMA), and 549 cm$^{-1}$ (PS-$d_8$)) and off-resonance conditions. Interestingly, the conductance is temperature-independent under cavity VSC of the $\delta$(Ar. CH(D)) modes of polymers, while off-resonance cavities showed a thermal activation (Fig.4C, Fig.S32). The path length and temperature-independent extraordinary conductance under cavity VSC of the $\delta$(Ar. CH(D)) modes suggest the possibility of a ballistic-like transport [3,4] or an insulator-to-metal transition [46]. The thermal activation under the VSC of non-$\delta$(Ar. CH(D)) modes and the off-resonance cavities point to a diffusive transport mechanism. However, more experiments are required to confirm the nature of transport.

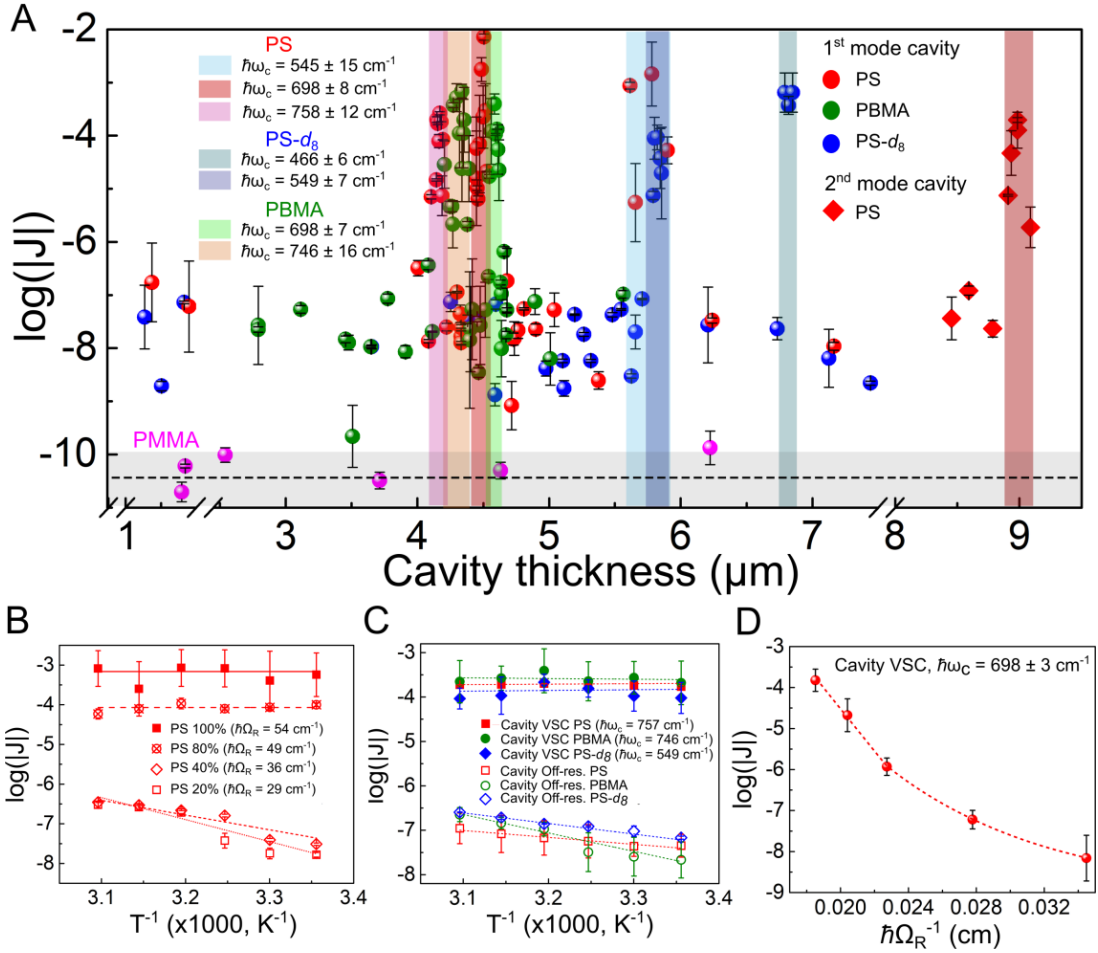

**Figure 4. Cavity thickness and temperature independent electrical conductance under cavity VSC of $\delta(Ar.\ CH(D))$ transition**. (A) The plot of current density as a function of the path length of cavities containing PS (red spheres and diamonds), PS-$d_8$ (blue spheres), PBMA (green spheres), and PMMA (pink spheres). The shaded region shows the path length of cavities corresponding to the cavity VSC of the different $\delta(Ar.\ CH(D))$ transition of PS, PS-$d_8$, PBMA and PMMA as labeled in the plot. The spheres and diamonds indicate the coupled or off-resonance condition with the cavity's first and second modes, respectively. (B) The plot showing the temperature dependence of log(|J|) under the cavity VSC of PS ($\hbar\omega_m$ = 698 cm$^{-1}$; $\hbar\omega_c$ = 698±3 cm$^{-1}$) for varying $\hbar\Omega_R$. (C) The plot showing the temperature dependence of log(|J|) under the cavity VSC of PS (solid red square; $\hbar\omega_c$ = 757 cm$^{-1}$), PS-$d_8$ (solid blue diamond; $\hbar\omega_c$ = 549 cm$^{-1}$) and PBMA (solid green circle; $\hbar\omega_c$ = 746 cm$^{-1}$) and off-resonance conditions (empty square, diamond and circle). (D) The plot of log(|J|) against $(\hbar\Omega_R)^{-1}$.

We introduce a theoretical model providing a qualitative explanation to understand the experimental findings. We use a phonon-assisted hopping model where electronic orbitals are connected together by the absorption and emission of phonons dressed by the cavity vacuum field. While phonons are ill-defined in strongly disordered media such as polymer films [47], the disorder is irrelevant on the long wavelengths scale ~10 µm of VSC. Strongly coupled vibrations can thus be effectively described by phonons propagating with well-defined in-plane momentum and polarization directions, which are then averaged over to characterize the transport properties.

The fields are decomposed over a set of in-plane wave vectors $k_\parallel$ and discrete modes $n > 0$ along the cavity axis. We consider nearest-neighbor electron hopping along the cavity axis between $N$ molecular orbitals with regularly spaced energies $\omega_j$ within a Gaussian profile of width W. The microscopic Hamiltonian ($\hbar = 1$) reads $H = H_{el} + H_m + H_{pol}$ (see supplemental material), with $H_{el} = \sum_j \omega_j c_j^\dagger c_j$ the bare electron Hamiltonian [$c_j$ annihilates an electron in the orbital localized on site $j$],

$H_m = \sum_{k_\parallel, n \neq 1,2} \omega_m b_{k_\parallel,n}^\dagger b_{k_\parallel,n} + \sum_{k_\parallel, n \neq 1,2, j} g_{k_\parallel,n,j}(c_{j+1}^\dagger c_j + h.c.)(b_{k_\parallel,n} + b_{-k_\parallel,n}^\dagger)$, and

$H_{pol} = \sum_{k_\parallel, n=1,2, \lambda} \omega_{k_\parallel,\lambda,n} p_{k_\parallel,\lambda,n}^\dagger p_{k_\parallel,\lambda,n} + \sum_{k_\parallel, n=1,2, \lambda, j} g_{k_\parallel,n,j} X_{k_\parallel,\lambda,n}(c_{j+1}^\dagger c_j + h.c.)(p_{k_\parallel,\lambda,n} + p_{-k_\parallel,\lambda,n}^\dagger)$.

$H_m$ describes hopping mediated by the bare phonon modes $n \neq 1,2$ that are unaffected by the cavity [$b_{k_\parallel,n}$ annihilates a dispersionless phonon with energy $\omega_m$ in the mode $(k_\parallel, n)$]. $H_{pol}$ is the polariton contribution, [$p_{k_\parallel,\lambda,n}$ annihilates an upper ($\lambda = UP$) or lower ($\lambda = LP$) polariton with energy $\omega_{k_\parallel,\lambda,n}$], which are linear superpositions of the phonon mode $n$ and the first transverse magnetic mode of the cavity, with in-plane polarization for $n = 1$ or along the cavity axis for $n = 2$. $X_{k_\parallel,\lambda,n}$ is the phonon weight of the polaritons and $g_{k_\parallel,n,j}$ the electron-phonon coupling strength. $\omega_{k_\parallel,\lambda,n}$ depends on the phonon-photon coupling strength $\Omega_0$ at normal incidence ($k_\parallel = 0$), which is connected to the Rabi splitting as $\Omega_0 = \pi \Omega_R/2$. The polariton dispersion averaged over the photon polarization directions is shown in Fig.5A and is independent of the coupled phonon mode $n$ (doubly degenerate spectrum), consistently with the transmission spectra.

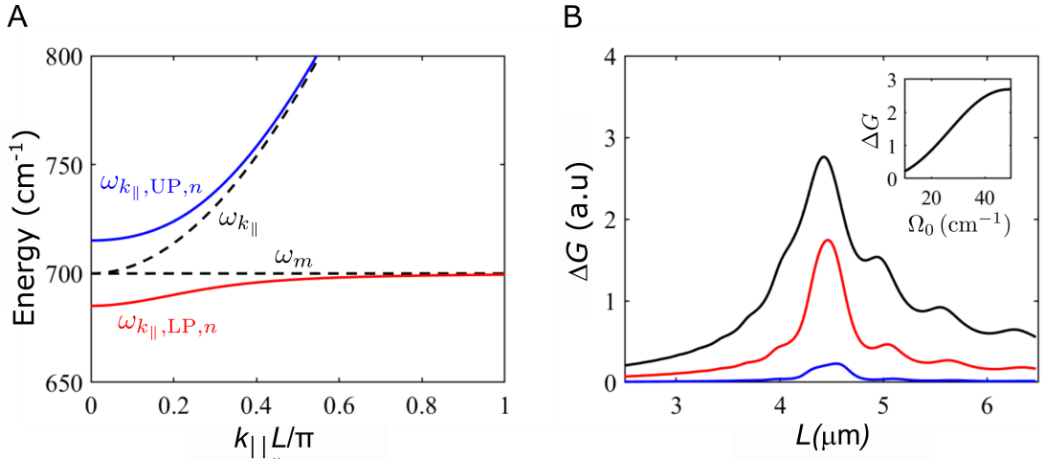

**Figure 5. Theoretical model.** (A) LP and UP polariton dispersion (red and blue solid curves) under resonant VSC ($L=L_{res}=$ 4.5 $\mu m$, $\Omega_0=47.1$ $cm^{-1}$) of the mode at $\omega_m=$ 698 $cm^{-1}$ and the first cavity mode $\omega_{k\parallel}$ (black dashed lines). $\omega_{k\parallel = 0}$ is denoted as $\omega_c$ in the main text. The dispersion is averaged over the photon polarization directions. (B) Conductance enhancement versus cavity length $L$ for different phonon-photon coupling strengths (blue: $\Omega_0 = 15.7 cm^{-1}$, red: $\Omega_0 = 31.4 cm^{-1}$, black: $\Omega_0 = 47.1 cm^{-1}$). Inset: Conductance enhancement at $L = L_{res} = 4.5 \mu m$ as a function of $\Omega_0$. Parameters: $T = 298K$, $\delta\omega = 60 cm^{-1}$, $W = 800 cm^{-1}$.

The averaged bulk conductance of the film reads $G = \sum_{n=1,2} G_n + \sum_{n \neq 1,2} G_{0n}$, with $G_n$ ($G_{0n}$) the contribution of the polariton (phonon) mode $n$ with (without) VSC. We find that the conductance enhancement $\Delta G = \sum_{n=1,2}(G_n - G_{0n})$ associated to the mode at 698 $cm^{-1}$ (Fig.5B) is peaked in the vicinity of the resonant cavity length $L_{res} \approx 4.5 \mu m$ at normal incidence and grows with $\Omega_0$, consistently with the experiments. This effect occurs for a broad range of W and is also present for the other modes of PS at 541 cm$^{-1}$ and 758 cm$^{-1}$ (see Supporting Information).

The conductance enhancement originates mostly from the LP, which lowers the activation energy $\omega$ of the transport processes $\propto \exp(-\omega/kT)$. Secondly, we find that the resonant feature is a finite-size effect as it only occurs for finite spacings $\delta\omega \sim W/N$ between two consecutive orbitals. Transport is enhanced by VSC when two orbitals can be efficiently interconnected by polaritons but not by bare phonons. The effect vanishes for $N \to \infty$, where orbitals can be interconnected by any excitation. The conductance enhancement occurs for $N \sim 10 - 100$, which corresponds to the ratio $L/l$, with $l \sim 100 nm$ the typical size of the polymers. The temperature dependence of the conductance is more challenging to elucidate since $G_{0n}$ also depends on $T$. Nevertheless, we show in the Supporting Information that the $T$-dependence is strongly affected under VSC.

Remarkably, we show for the first time that one can transform the intrinsically non-conducting amorphous polymers (PS, PS-$d_8$, and PBMA) into a conductor by simply placing them between two mirrors at the proper distance. The conductivity of PS under VSC (c.a. 9 Sm$^{-1}$) is comparable to graphene-doped polystyrene [48]. Thus, VSC provides a handle to tune the electron/hole transport through molecular materials, offering new possibilities for organic electronics and semiconductor technology. The device preparation is straightforward, providing another platform for lossless long-range transport in molecular devices. All the polymers we studied are intrinsic non-conductors; therefore, the total energy barrier ($\Phi$) for electrical conductance is significantly high [49]. Under VSC, however, the conductance is dramatically enhanced, indicating that $\Phi$ has an inverse relationship with $\hbar\Omega_R$ (Fig.4D). It will be interesting to investigate how $\Phi$ can undergo such a drastic reduction under VSC, which is beyond predictions of our current theoretical model for bulk conductance. Further experiments and theoretical analyses, such as molecular dynamics simulations, will be needed to understand the mechanistic details of this remarkable transport phenomena observed under VSC.

**Acknowledgments:** We thank the Department of Inorganic and Physical Chemistry, Center for Nanoscience and Engineering, Dr. P. Rajamalli, and Prof. S. Ramakrishnan for providing access to instrumentation facilities. We thank Prof. S. Vasudevan and Prof. T. W. Ebbesen for gifting us deuterated polystyrene and helpful discussions. We thank Ms. Anupama (visiting student from CUSAT, Kochi) for her help in preparing PBMA cavities. SB, UR, and KSM thank DST-INSPIRE and the Prime Minister's Research Fellowship (PMRF) for the Ph.D. fellowship. SK thanks UGC for the research fellowship. RV thanks the European Research Executive Agency for funding under the European Union Horizon TMA MSCA Postdoctoral Fellowships – European Fellowships action. GC thanks SERB for fellowship. AT and VK thanks IISc for a start-up grant and DST-SERB for research funding. AT thanks Infosys Young Investigator Award, and VK thanks DST-INSPIRE Faculty Fellowship. We thank Prof. S. Sampath, Prof. E. Arunan, Prof. P. K. Das, Prof. Sai G Ramesh, Prof. Srihari Keshavamurthy, Dr. Thibault Chervy, and Dr. Eloise Devaux for their helpful discussion.

**Funding:** Horizon TMA MSCA Postdoctoral Fellowship – European Fellowship action (Grant Agreement No 101068621) for RV. IISc start-up grant for VK (SG/MHRD-19-0034 and SR/MHRD-19-0028) and AT (IE/CARE-21-0321 and SR/MHRD-20-0039), SERB-CRG grant for VK (CRG/2020/002302) and AT (CRG/2021/002396), Infosys Young Investigator Award for AT (FG/OTHR-21-1050), INSPIRE Faculty fellowship for VK (DST/INSPIRE/04/2018/002983)

**Author contributions:** All the authors contributed to the results presented in the paper. AT and VK conceived the idea and supervised the project. SB took the lead in preparing and characterizing all the strongly coupled samples with the help of GC. SK took the lead in the electrical measurements. UR and VK build the EGaIn setup. KSM and RV did the Transfer-Matrix simulations. DH developed the theory model and did the calculations. All the authors analyzed the data discussed together and wrote the manuscript.

**Competing interests:** AT, VK, SK, and SB are declared inventors on the patent application submitted by the Indian Institute of Science (Application No: 202341014975)

**Data and materials availability:** All data are in the main text or the Supporting Information. Additional details can be provided upon reasonable request.

**Supporting Information**

Materials and Methods

Figures S1-S148

Table S1, S2, S3, S4 and S5